\documentclass[aps,prb,twocolumn,superscriptaddress]{revtex4-2}
\usepackage[utf8]{inputenc}
\usepackage{graphicx}
\usepackage{booktabs}
\usepackage{wrapfig}
\usepackage{chemformula}
\usepackage{xcolor}
\usepackage{comment}
\usepackage{placeins}

\begin{document}

\title{Oxygen-Mediated Phase Evolution in Sputtered Cu-W-O: Insights into Surface Chemistry Variability}

\author{José Montero-Amenedo}
\email{[jose.montero@dci.uhu.es](mailto:jose.montero@dci.uhu.es)}
\affiliation{Departamento de Ciencias Integradas, Universidad de Huelva, 21071 Huelva, Spain}

\begin{abstract}
Thin films of Cu–W–O ternary compounds were fabricated via DC magnetron co-sputtering from Cu and W metallic targets under controlled oxygen partial pressures, followed by thermal annealing. Low-oxygen conditions favored the formation of a single \ch{CuWO4} phase, whereas higher oxygen levels produced a mixture of \ch{CuWO4} and \ch{Cu3WO6}.
Structural and optical properties were investigated by X-ray diffraction (XRD) and spectrophotometry, revealing phase coexistence and changes in preferential orientation depending on the deposition conditions. A detailed and carefully validated X-ray photoelectron spectroscopy (XPS) analysis provides insight into the surface chemical environment of Cu and W, indicating the presence of compositional inhomogeneities and surface–bulk differences associated with Cu migration and segregation.
While the W 4f core levels remain remarkably stable across all tested oxygen partial pressures, a systematic shift is observed in the Cu 2p$_{3/2}$ binding energy. Wagner plot analysis confirms that this displacement is dominated by initial-state effects, reflecting modifications of the Cu ground-state electronic structure and Cu–O–W hybridization rather than changes in final-state screening.
Our findings demonstrate that sputtered Cu–W–O films, even when nominally identified as \ch{CuWO4}, can exhibit substantially different structural and electronic states depending on synthesis conditions, highlighting the need for rigorous characterization to ensure reproducibility in ternary oxide research.
\end{abstract}

\keywords{Sputtering, \ch{CuWO4}, \ch{Cu3WO6}, Wagner plot, Surface chemistry, Initial-state effects}

\maketitle

\section{Introduction}

Over the past decade, interest in \ch{CuWO4} has increased significantly—with annual publications growing nearly eight-fold from 2015 to 2025 \cite{GoogleScholarCuWO4}. This surge is driven by the exploration of ternary metal oxides as a prominent alternative to conventional binary systems such as \ch{WO3} and \ch{CuO}. The combination of multiple cations facilitates increased flexibility in tuning functional properties and enables emergent behaviors not accessible in simpler oxides \cite{Sivakumar2025_synergy_transition_ternary_oxides, Liu2025TungstenPhotocatalysts}.

\ch{CuWO4} has been widely investigated for applications such as photocatalytic pollutant degradation \cite{Liu2025TungstenPhotocatalysts} and photoelectrochemical water splitting \cite{Arunraj2025_photoanodes_record}. Additional applications include chemical sensing \cite{Akila2026CuWO4Sensing} and wastewater treatment via adsorption processes \cite{Li2018HollowCuWO4_adsorption}. However, the reported functional performance of \ch{CuWO4} varies significantly across studies, even under seemingly comparable conditions \cite{Arunraj2025_photoanodes_record,Liu2025TungstenPhotocatalysts}.

Such variability is often attributed to differences in morphology and microstructure arising from specific synthesis methods \cite{Liu2025TungstenPhotocatalysts, Li2018HollowCuWO4_adsorption}. While these factors are undoubtedly significant, they do not fully account for the large discrepancies frequently observed in experimental data.
In fact, literature reviews \cite{Roy_2018_variability_gap, Rajeshwar2018CopperOxideReview} indicate that ternary oxides generally exhibit a much higher degree of variability in their inherent electronic properties—most notably the reported optical bandgap—compared to their binary counterparts. Furthermore, recent computational studies demonstrate that specific surface terminations in \ch{CuWO4}, such as the (010) and (101) planes, fundamentally dictate adsorption energetics and water-splitting efficiency \cite{Chu2023Redox, Chu2024Adsorption, Chu2025Mechanism}.
These findings suggest that surface composition, phase coexistence, and local electronic structure are frequently overlooked and remain insufficiently understood, yet they likely constitute the primary drivers of the observed performance variations.

In this work, we demonstrate that sputtered Cu–W–O thin films—which might otherwise be nominally labeled as \ch{CuWO4} if their inherent phase-heterogeneity were overlooked—exhibit substantial variations in structural, compositional, and electronic properties depending on the deposition conditions. These variations include the coexistence of \ch{CuWO4} with amorphous \ch{CuO} phases, the formation of Cu-rich compounds such as \ch{Cu3WO6}, changes in preferential crystallographic orientation, and significant differences between bulk and surface composition indicative of Cu migration and segregation.
A detailed X-ray photoelectron spectroscopy (XPS) analysis, performed following rigorous fitting and validation procedures, reveals pronounced variations in the local chemical environment of Cu, while W remains comparatively stable. These findings highlight the sensitivity of the Cu sublattice to processing conditions, which may play a key role in determining surface properties.
Overall, our results show that \ch{CuWO4} sputtered thin films can exhibit significant deviations from the ideal composition and structure. While such effects are commonly observed in sputtered oxide films, they can be further exacerbated in multication systems like \ch{CuWO4}, where the presence of multiple cations introduces additional degrees of freedom in phase formation, diffusion, and defect chemistry. While this flexibility may be advantageous for tuning functional properties, it also increases the sensitivity of the material to synthesis conditions. Consequently, variations in surface and chemical states may contribute significantly to the large dispersion of functional properties reported in the literature.

\section{Experimental}

\begin{table}[ht]
\centering
\begin{tabular}{|p{0.35\linewidth}|p{0.6\linewidth}|}
\hline
\textbf{Parameter} & \textbf{Details} \\
\hline
Sputtering System & Balzers UTT 400. \\
\hline
Targets & Cu and W (ø 5.8 cm, 99.995 \%). \\
\hline
Mode & DC, constant power $P = 200 \, \text{W}$ (W), $250 \, \text{W}$ (Cu). \\
\hline
Base Pressure & $10^{-7}$ mbar. \\
\hline
Working Gas & Argon (99.998 \% pure), Flow rate: $\phi_{\ch{Ar}} = 50$ sccm. \\
\hline
Working Pressure & $4 \times 10^{-2}$ mbar. \\
\hline
Reactive Gas & \ch{O2}, flow rate $\phi_{\ch{O2}}$ between 5.0 sccm and 30.0 sccm. \\
\hline
Film Thickness & $\sim$ 300 nm. \\
\hline
Substrate & Soda-lime glass. \\
\hline
\end{tabular}
\caption{Sputtering parameters used for the deposition of precursor samples.}
\label{tab:sputtering_parameters}
\end{table}

Thin films consisting of Cu-W-O ternary compounds, including \ch{CuWO4} and \ch{Cu3WO6}, were prepared via a two-step synthesis process: deposition of Cu–W–O precursor films by DC magnetron sputtering onto glass substrates, without intentional substrate heating, followed by annealing of the as-sputtered films in air at 500$^\circ$C for 20 minutes.

The deposition was carried out in a Balzers UTT 400 sputtering system using two metallic targets: Cu (diameter: 5.8 cm, purity: 99.995 \%) and W (diameter: 5.8 cm, purity: 99.995 \%). The process was as follows: the chamber was first evacuated to a base pressure of $10^{-7}$ mbar, then high-purity argon (flow $\phi_{Ar} = 50$ sccm, 99.998\%) was introduced, raising the chamber pressure, which was regulated at $4\times10^{-2}$ mbar using a throttle valve between the vacuum line and the deposition chamber. Constant powers of 200 W and 250 W were applied to the W  and Cu  targets, respectively, based on prior experience in the preparation of copper oxide \cite{Thyr2022ZnOCuOxides, montero2018_CuOx} and tungsten oxide \cite{Atak2020N-doped,Atak2021Judicious} thin films in this system.

Before deposition, the targets were pre-sputtered in pure Ar for 5 minutes to minimize undesired hysteresis effects \cite{Safi2000ReactiveSputtering}, while the substrate was protected by a shutter. After pre-sputtering, \ch{O2} was introduced into the chamber, and once the plasma stabilized, the shutter was opened to allow film growth. The oxygen partial pressure, $P_{O_2}$, was calculated from the pressure increase after introducing \ch{O2}. Samples were deposited at \ch{O2} flow rates ($\phi_{O_2}$) ranging from 5.0 sccm to 20.0 sccm in 2.5 sccm increments.

During deposition, the sputtering discharge current $I$ was monitored for each target. Deposition time was adjusted to achieve a film thickness of approximately 300 nm, measured using the step height between the substrate and the film surface with a Bruker Dektak profilometer. Table \ref{tab:sputtering_parameters} summarizes the sputtering parameters used for precursor sample deposition.

Additional characterization included grazing-incidence X-ray diffraction (GIXRD) using a Siemens D5000 diffractometer with Cu K$_\alpha$ radiation at a 1$^\circ$ angle of incidence.  

Optical characterization, including transmittance ($T$) and reflectance ($R$) measurements, was performed with a Perkin--Elmer Lambda 900 spectrophotometer equipped with a calibrated integrating sphere.

Surface composition and oxidation states were analyzed by X-ray photoelectron spectroscopy (XPS) using an ULVAC PHI Quantera II spectrometer equipped with a monochromatic Al K$\alpha$ source. High-resolution spectra of the Cu 2p, O 1s, W 4f, and C 1s core levels were acquired at a pass energy of 55 eV with an energy step size of 0.2 eV. Survey spectra were recorded at a pass energy of 224 eV and a step size of 0.8 eV. Charge compensation was applied using the instrument's dual-beam electron/ion neutralizer.
XPS spectra were analyzed using the commercial software CasaXPS \cite{Fairley2021CasaXPS}. Peak fitting was performed using a Shirley background, and spectral components were modeled with Gaussian–Lorentzian line shapes, denoted GL(30) in CasaXPS, corresponding to a 30\% Lorentzian contribution. The peak-fitting procedures and spectral constraints were implemented in accordance with current best practices for XPS data analysis \cite{major2020practical}.
Energy calibration was carried out using the adventitious carbon C 1s peak at 284.8~eV. Elemental quantification was performed using relative sensitivity factors (RSFs) imported from the library of the MultiPak software package (version 9.8.0.19, ULVAC-PHI, Inc.) \cite{PHI_MultiPak_9_8_0_19}, which are specific to the PHI Quantera II instrument.  
 
 \section{Results and Discussion}

 \subsection{Reactive sputtering considerations}

 \begin{figure}
\includegraphics[width=0.45\textwidth]{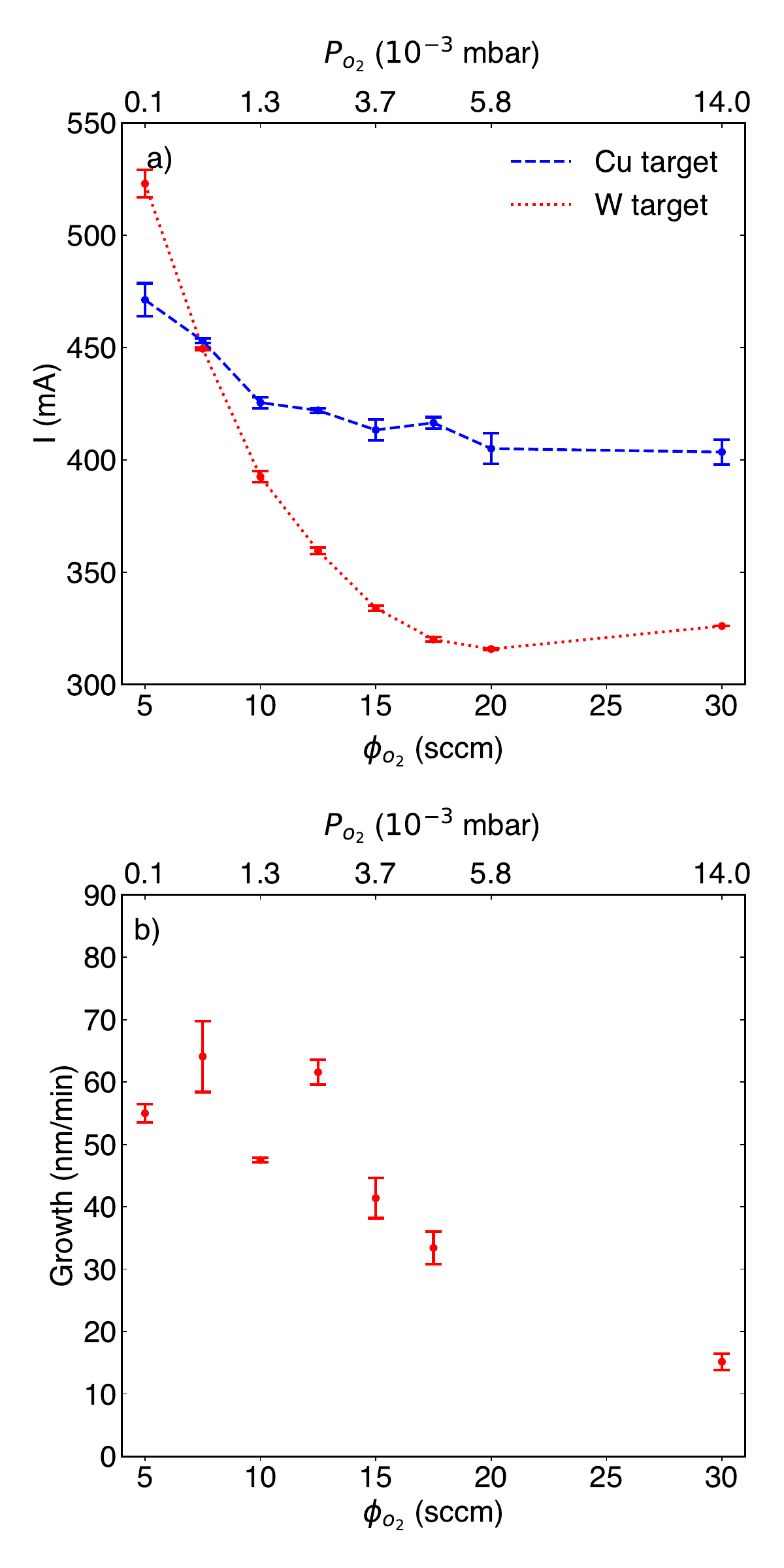}
\caption{Sputtering discharge current $I$ (a) for the Cu and W targets—dashed and dotted lines are intended as guides to the eye—and the films’ growth rate (b), as a function of oxygen partial pressure $P_{O_2}$ or oxygen flow $\phi_{O_2}$. Both panels include datasets with error bars.}
\label{fig1}
\end{figure}

The co-sputtering of the Cu and W targets for the fabrication of precursor samples was performed at fixed power. As typical in reactive sputtering systems, variations in the discharge current $I$ and the films’ growth rate were observed as a function of the reactive gas inlet, in this case oxygen \cite{Safi2000ReactiveSputtering}.
Figure \ref{fig1}(a) shows the discharge current $I$ for the Cu and W targets as a function of $P_{O_2}$ and $\phi_{O_2}$, while Figure \ref{fig1}(b) presents the corresponding growth rate in nanometers per minute. The curve in Figure \ref{fig1}(b) exhibits the typical behavior of reactive sputtering: the transition from metallic to reactive mode as $\phi_{O_2}$ (and thus $P_{O_2}$) increases, which results in a reduction of the sputtering yield \cite{Safi2000ReactiveSputtering}. In other words, target poisoning with increasing $P_{O_2}$ leads to a decrease in growth rate, from $55 \pm 2$ nm/min at $\phi_{O_2} = 5$ sccm to $15 \pm 1$ nm/min at $\phi_{O_2} = 20$ sccm. 
The growth-rate information shown in Fig.~\ref{fig1}(b) was subsequently used to adjust the deposition time in order to achieve the targeted film thickness of approximately 300 nm.

The shift toward the reactive mode with increasing $\phi_{O_2}$ also causes a decrease in $I$ for both targets (Figure \ref{fig1} (a)). At $\phi_{O_2} < 7.5$ sccm, $I$ is higher for W than for Cu—for example, at $\phi_{O_2} = 5$ sccm, $I = 523 \pm 6$ mA for W and $471 \pm 7$ mA for Cu. Above 7.5 sccm, the situation reverses: at $\phi_{O_2} = 20$ sccm, $I = 326 \pm 0$ mA for W and $403 \pm 5$ mA for Cu.

Although the discharge current trend cannot be interpreted as a direct measure of the sputtering yields of W and Cu under reactive conditions, the stronger current reduction observed for the W target with increasing $\phi_{O_2}$ indicates a more pronounced modification of the plasma–target interaction compared to the Cu target. Consequently, a sharper decrease in the sputtering rate is expected for the W target than for the Cu target when transitioning from low to high oxygen conditions.
Therefore, Fig.~\ref{fig1}(a) suggests that changes in the oxygen flow during deposition not only influence the overall sputtering rate, as shown in Fig.~\ref{fig1}(b), but may also lead to variations in the Cu/W ratio in the resulting films, particularly an increase in Cu/W with increasing oxygen flow.

Beyond the intrinsic properties of the target materials, the different discharge behavior observed in Fig.~\ref{fig1}(a) may also be influenced by geometric factors of the co-sputtering configuration (e.g., magnetic field distribution, erosion track area, and the relative positioning of the targets with respect to the oxygen inlet), which can modify plasma parameters and ion flux independently of intrinsic material properties.

\subsection{Structural and Optical Properties}

\begin{figure} 
\centering
\includegraphics[width=0.45\textwidth]{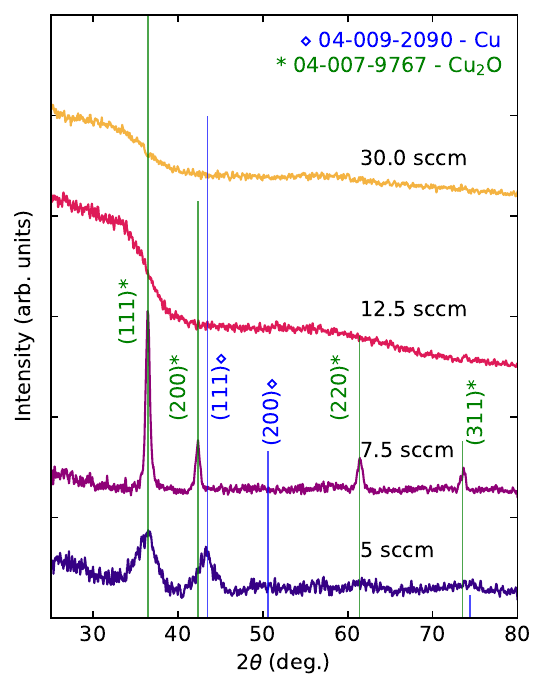}
\caption{X-ray diffraction patterns corresponding to the as-sputtered (not annealed) Cu-W-O samples at different oxygen flows: 5, 7.5, 12.5 and 30 sccm. Drop lines correspond to XRD peaks corresponding to metallic Cu ($\diamond$) and \ch{Cu2O} ($\ast$) according to ICDD cards No.~04-009-2090 and No.~04-007-9767.}
\label{fig2}
\end{figure}

Figure \ref{fig2} shows the XRD patterns corresponding to selected as-deposited samples (prior to the annealing process). Samples deposited at $\phi_{O_2} = 5$ sccm exhibit characteristic peaks associated with \ch{Cu2O} and metallic Cu--International Centre for Diffraction Data (ICDD) cards No.~04-007-9767 and 04-009-2090. In contrast, at $\phi_{O_2} = 7.5$ sccm, only peaks corresponding to \ch{Cu2O} are observed. No tungsten-containing phases are detected by XRD under any deposition conditions. For oxygen flow rates above $\phi_{O_2} > 7.5$ sccm, no crystalline features are observed.

\begin{figure*} 
\includegraphics[width=0.99\textwidth]{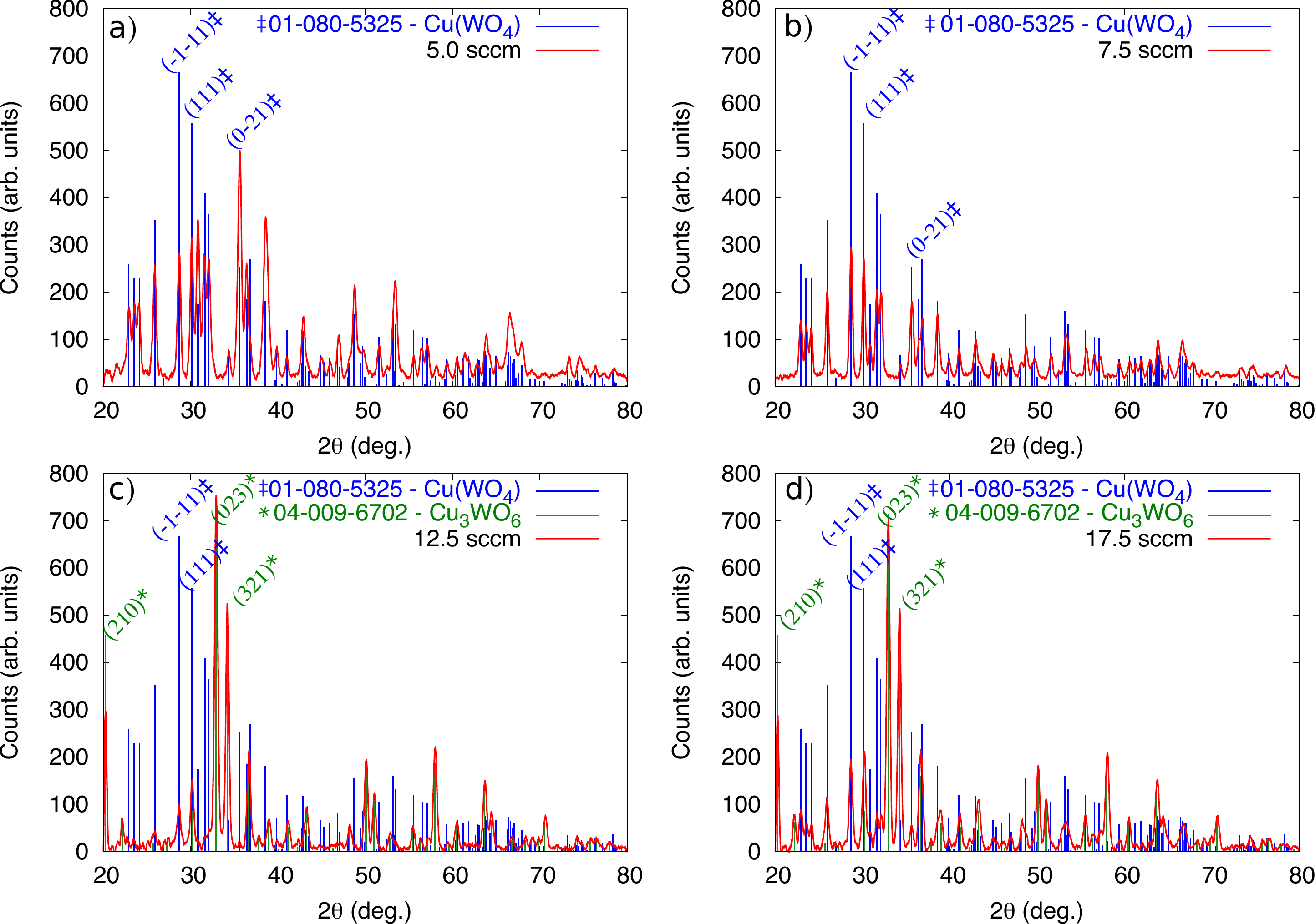}
\caption{Grazing incidence X-ray diffraction (GIXRD) patterns for samples deposited at oxygen flows of 5.0 (a), 7.5 (b), 12.5 (c), and 17.5~sccm (d). Vertical bars indicate reference positions from ICDD cards No.~01-080-5325 (\ch{CuWO4}) and No.~04-009-6702 (\ch{Cu3WO6}), with selected Bragg peaks labeled with $\ddagger$ and $\ast$, respectively.}
\label{fig4}
\end{figure*}

In order to promote the formation of the Cu-W-O ternary phases, the sputtered precursor films were subjected to an annealing process in air. During this process, the temperature was ramped from ambient to 500 $^\circ$C over 50 minutes, maintained at 500$^\circ$C for 20 minutes, and subsequently allowed to cool naturally to room temperature.

Representative X-ray diffractograms of annealed samples deposited at $\phi_{O_2} = 5.0$, 7.5, 12.5, and 17.5 sccm are presented in Figure \ref{fig2}, panels (a)–(d), respectively. Triclinic \ch{CuWO4} characteristic diffraction peaks, consistent with the ICDD card No.~01-080-5325, are observed in all samples irrespective of the deposition conditions.
For oxygen flow rates $\phi_{O_2} \leq 10$ sccm, only the crystalline \ch{CuWO4} phase is detected (Figure \ref{fig2} (a)–(b)). No secondary phases, including \ch{Cu2O}, \ch{CuO}, or \ch{WO3}, are observed by XRD. However, for $\phi_{O_2} > 10 $ sccm, in addition to triclinic \ch{CuWO4}, a cubic copper tungsten oxide \ch{Cu3WO6} phase \cite{gebert1969_Cu3WO6} is detected (ICDD card No.~04-009-6702), as shown in Figure \ref{fig2}c–d.

Regarding the texture of the \ch{CuWO4} phase, samples deposited at $\phi_{O_2} = 5$ sccm exhibit a marked preferential orientation, as evidenced by the enhanced intensity of the (0-21) Bragg peak (Figure \ref{fig2}a). In samples deposited at $\phi_{O_2} = 7.5$ sccm, the diffraction pattern displays intensity ratios consistent with the powder diffraction standard of \ch{CuWO4}, with the most intense Bragg peak corresponding to the (-1-11) plane, followed by the (111) plane (Figure \ref{fig2}b). A similar behavior is observed for samples deposited at $\phi_{O_2} > 7.5$ sccm, where no preferential orientation of the \ch{CuWO4} phase is detected (Figure \ref{fig2}c,d).

The change in preferred orientation of the \ch{CuWO4} phase may reflect a modification of the relative surface stabilities with increasing oxygen chemical potential, possibly associated with a change in surface termination. We have reported similar behavior in other oxide thin films grown by sputtering \cite{Montero2014SnO2Sb}, where variations in oxygen chemical potential alter the surface free energies and consequently the favored crystallographic orientation \cite{Korber2010preferential}. Crystallographic orientation may influence functional properties (e.g., photocatalytic or adsorption-related behavior) through several surface-dependent factors, such as the electronic band alignment \cite{Montero2014SnO2Sb}. In the particular case of \ch{CuWO4}, recent studies \cite{Chu2023Redox, Chu2024Adsorption, Chu2025Mechanism} have demonstrated that different surface terminations, such as the (010) and (101) planes, exhibit distinct water-adsorption and water-splitting capabilities.

As mentioned before, for $\phi_{O_2} > 7.5$ sccm, additional Bragg peaks corresponding to the Cu-rich \ch{Cu3WO6}  phase are observed. The intensity ratios of these peaks match those reported in the powder standard, with the most intense Bragg peak associated with the (023) plane, followed by the (321) plane, Fig. \ref{fig2} (c) and (d).

\begin{figure} 
\includegraphics[width=0.5\textwidth]{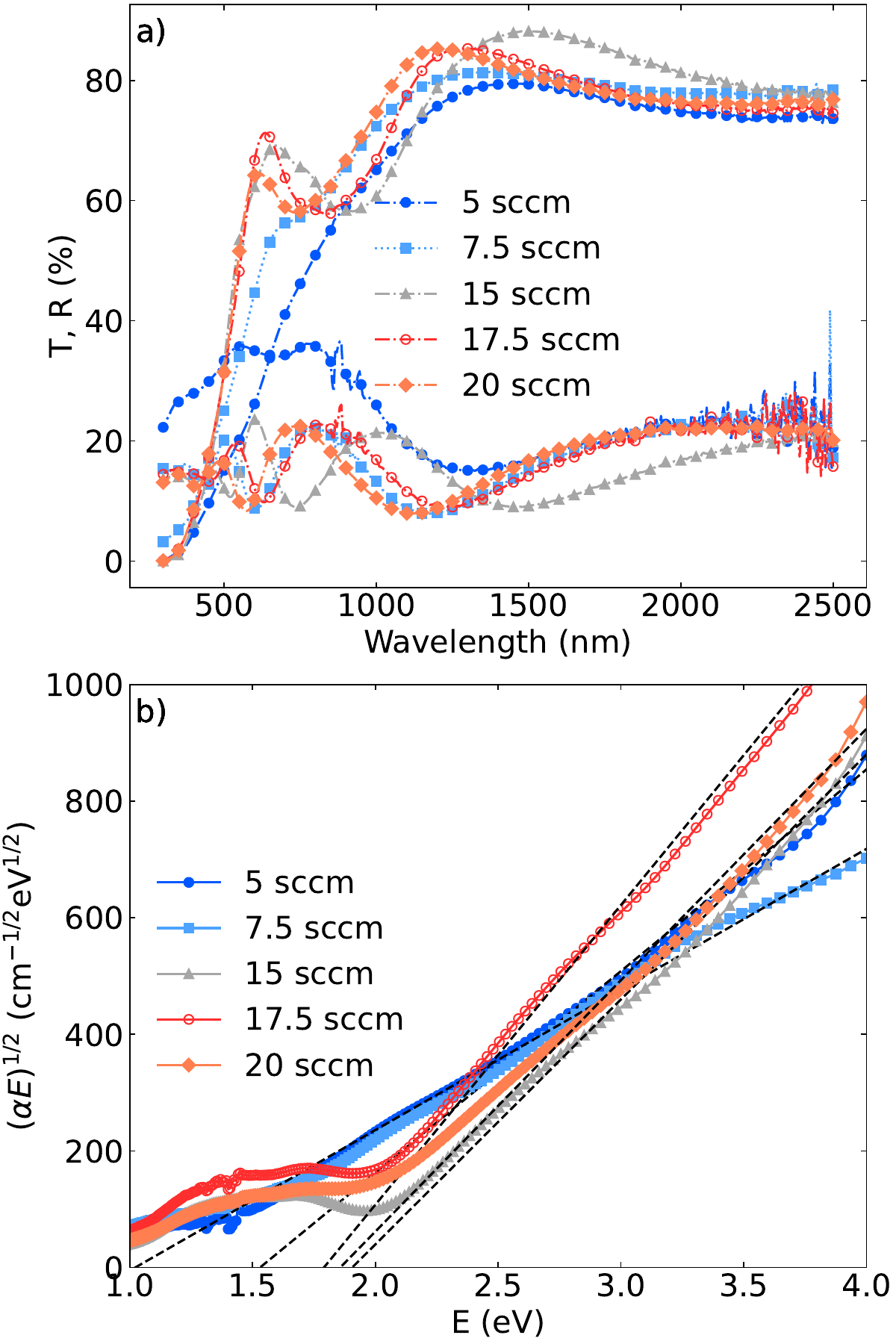}
\caption{Transmittance ($T$) and reflectance ($R$) spectra vs. wavelength (a); the $T$ curves correspond to the higher percentage values. Tauc plots for an indirect allowed transition ($(\alpha E)^{1/2}$ vs. $E$) for the same samples (b), where $\alpha$ is the absorption coefficient and $E$ is the photon energy. The optical bandgap is determined from the intercept of the linear fit (dashed lines) with the photon energy axis.}
\label{optics}
\end{figure}

All samples exhibit a yellowish-brown coloration visible to the naked eye, while remaining partially transparent in the visible range. Figure \ref{optics} (a) shows $T$ and $R$ as a function of wavelength for a series of annealed films deposited at various $\phi_{O_2}$ values. The fundamental absorption edge extends from the UV region into the visible spectrum for all samples, but particularly for the films deposited under low oxygen conditions, 5 and 7.5 sccm.

The absorption coefficient $\alpha$ was calculated from the measured $T$ and $R$ values using Hong’s formula \cite{Hong1989alpha}:

\begin{equation}
\alpha = -\frac{1}{d} \ln\left(\frac{T}{1 - R}\right),
\end{equation}

\noindent where $d$ is the film thickness.

The structural variations observed in the XRD patterns are expected to influence the optical properties of the films. To quantify this effect, the optical band gap was determined using the Tauc method by plotting $(\alpha E)^{1/2}$ as a function of photon energy $E$ (Figure \ref{optics}(b)), where the exponent $1/2$ corresponds to an indirect allowed transition \cite{Pankove1975}, which is typical for Cu–W–O ternary compounds \cite{Rajeshwar2018CopperOxideReview, Benko1982Cu3WO6}. Linear extrapolation of the high-energy region yields the band gap values summarized in Table~\ref{tab:xrd_bandgap}.

Table~\ref{tab:xrd_bandgap} lists the oxygen flow during deposition ($\phi_{\mathrm{O}_2}$), the XRD-derived structural characteristics before and after annealing, and the resulting optical parameters: the extracted band gap $E_g$ and the regression coefficient $r$ for the Tauc fit shown in Figure \ref{optics}(b).The linear region used for the Tauc analysis extends from  2.5 to 4.1 eV ($\sim$ 500–300 nm). In this range, the fits yield correlation coefficients close to $r = 0.99$ for most samples, indicating a good agreement with the indirect allowed transition model.

All samples, except those deposited at lower oxygen flows (5 and 7.5 sccm), exhibit an indirect allowed optical band gap of around 1.8–1.9 eV (Table~\ref{tab:xrd_bandgap}), consistent with values reported in the literature for \ch{CuWO4} ($\sim$ 1.8–2.4 eV) and \ch{Cu3WO6} ($\sim$ 1.85 eV), see Refs.~\cite{Benko1982Cu3WO6, Rajeshwar2018CopperOxideReview} and references therein. For samples deposited at 5 and 7.5 sccm, the measured indirect allowed band gaps of $\sim$ 1.0 and 1.5 eV suggest the presence of an amorphous CuO phase not detectable by XRD. Indeed, sputtered CuO thin films have been reported to exhibit an indirect allowed optical transition around 1.28 eV, as observed in our previous work \cite{montero2018_CuOx}. 
The presence of CuO with such a small band gap, combined with the structural disorder associated with its amorphous nature--which gives rise to Urbach tails \cite{Deo2026_UrbachCuO}--can account for the more pronounced sub-band-gap absorption tail extending into the visible region for films deposited at low oxygen flows.
By comparison, other possible binary oxides exhibit significantly larger band gaps: \ch{WO3} shows an indirect transition around 3.0 eV \cite{Atak2020N-doped}, while \ch{Cu2O} exhibits direct forbidden and direct allowed transitions at approximately 2.06 and 2.55 eV, respectively \cite{montero2018_CuOx}.

This interpretation pointing to the presence of an additional amorphous CuO phase in the samples deposited at low oxygen flow is further supported by the compositional trend observed for these films, where the Cu/W ratio increases beyond the stoichiometry expected for ternary Cu–W–O compounds, as revealed by XPS and discussed in detail below.

\begin{table}
\centering
\caption{Phase evolution before and after annealing determined by XRD, alongside the corresponding optical bandgap values $E_g$ (for an indirect allowed transition) and the regression coefficient ($r$) from the Tauc plot linear fits (Fig.~\ref{optics} (b)), as a function of oxygen flow rate ($\phi_{\mathrm{O}_2}$).}
\label{tab:xrd_bandgap}
\footnotesize
\setlength{\tabcolsep}{4pt}
\begin{tabular}{ccccc}
\hline
$\phi_{\mathrm{O}_2}$ (sccm) & Unannealed & Annealed & $E_g$ (eV) & $r$ \\
\hline
5.0   & Cu + \ch{Cu2O}         & \ch{CuWO4}                 & 1.5 & 0.99 \\
7.5   & \ch{Cu2O}              & \ch{CuWO4}                 & 1.0 & 0.99 \\
12.5  & Amorph.                & {\ch{CuWO4 + Cu3WO6}}       & 1.9 & 0.97 \\
15.0  & Amorph.                & \ch{CuWO4 + Cu3WO6}       & 1.9 & 0.99\\
17.5  & Amorph.                & \ch{CuWO4 + Cu3WO6}        & 1.8 & 0.99 \\
20.0  & Amorph.                & \ch{CuWO4 + Cu3WO6}        & 1.9 & 0.99 \\
\hline

\end{tabular}
\end{table}

\subsection{XPS: Surface Composition and Chemical States}

\begin{figure*}

    \centering
    \includegraphics[width=0.7\textwidth]{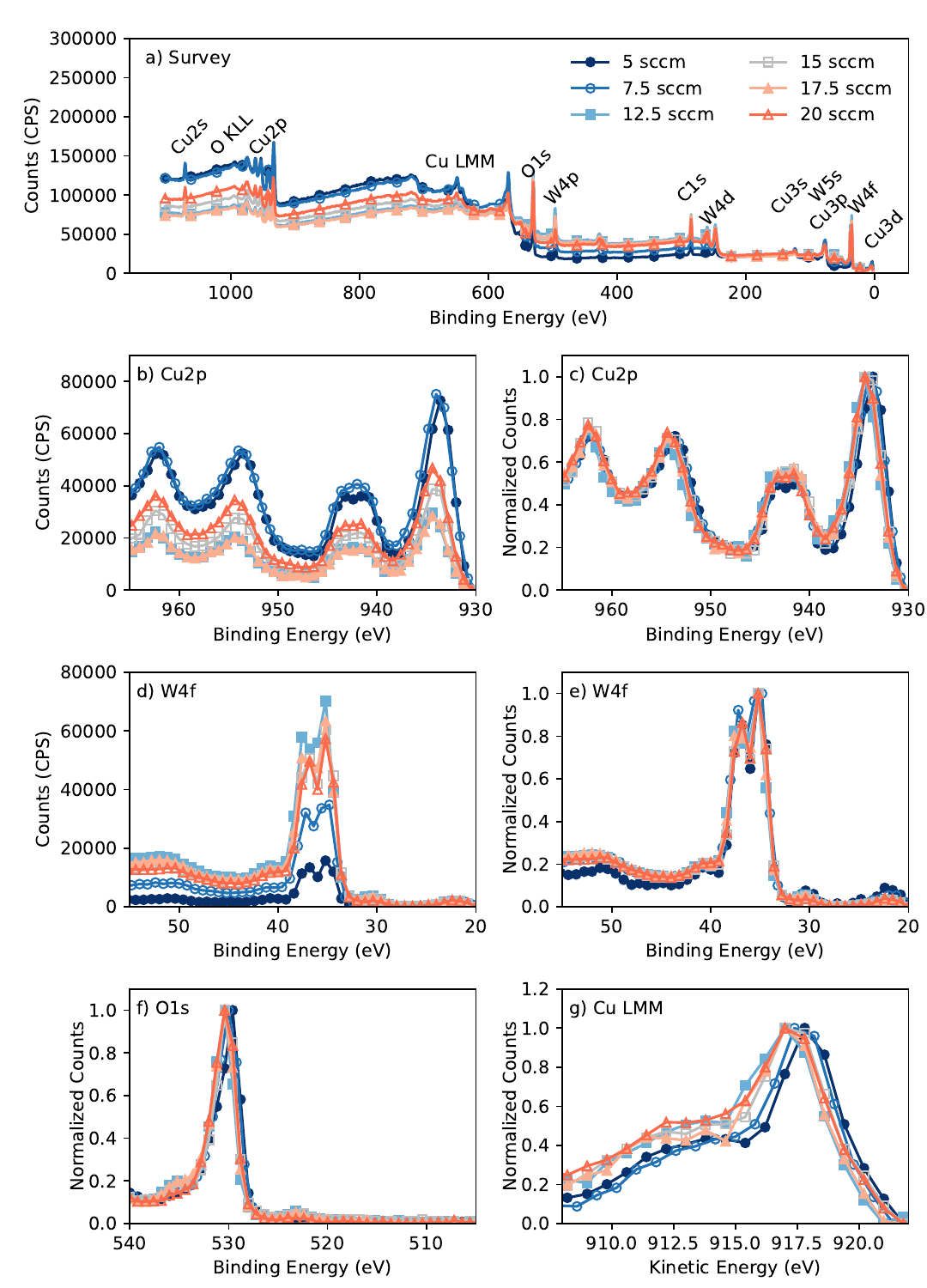}
\caption{Evolution of the surface chemical states with oxygen flow rate, derived from survey spectra. (a) Full-range survey spectra. Cu~2p and W~4f core-levels, displayed with a baseline offset (b, d)---where each spectrum is shifted so that its minimum intensity is zero---and normalized (c, e). Normalized O~1s core-level region (f).  Cu~L$_3$M$_{45}$M$_{45}$ Auger spectra (g).}
    \label{fig:survey}
\end{figure*}

X-ray photoelectron spectroscopy (XPS) was employed to investigate the near-surface chemical composition and oxidation states of the films as a function of oxygen flow. The analysis is structured as follows: we first examine the Cu~2p, W~4f, and O~1s core-level regions to identify elemental trends \textit{(i)}, complemented by Cu~LMM Auger analysis and a Wagner plot to distinguish between initial- and final-state chemical effects \textit{(ii)}. This is followed by an assessment of the surface composition and phase interpretation \textit{(iii)}. We then compare these surface-sensitive results with bulk measurements to discuss the role of copper migration and phase segregation \textit{(iv)}. Finally, we provide a detailed interpretation of the surface chemical states through high-resolution peak fitting \textit{(v)}.

\begin{figure}
    \centering
    \includegraphics[width=0.5\textwidth]{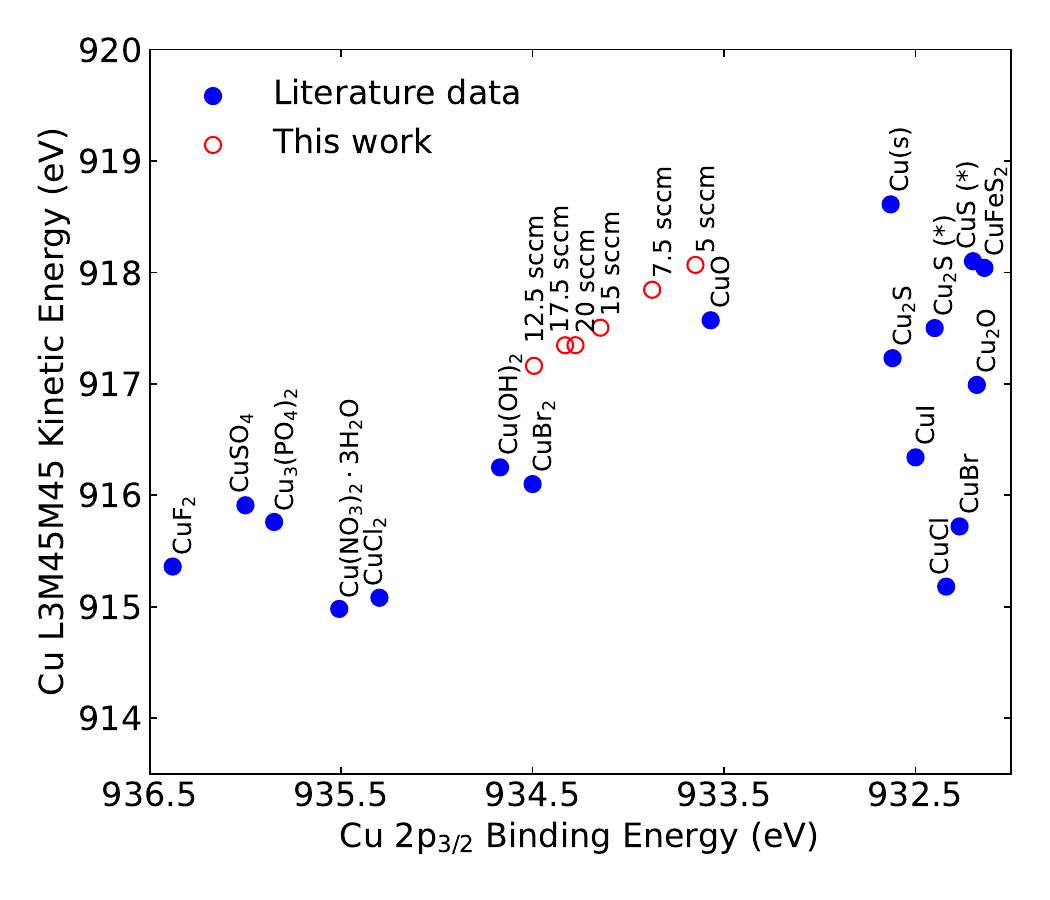}
    \caption{Wagner plot showing Cu 2p binding energy versus Cu LMM kinetic energy for literature data (blue open circles) compiled by Moretti and Beck~\cite{Moretti_Beck_2019} and the present work (red filled circles). The compilation is primarily based on measurements by Biesinger~\cite{Biesinger2017CuXPS}, with additional data for \ch{Cu2S} and \ch{CuS} from Perry and Taylor~\cite{PerryTaylor1986_Auger_Cu2S_CuS}, denoted by an asterisk ($\ast$). Labels indicate chemical formulas for literature points and deposition conditions for samples from this study.}
    \label{fig:wagner}
\end{figure}

\medskip
\noindent\textit{(i) Core-level evolution and elemental trends.} 
\medskip

Figure~\ref{fig:survey}(a) shows XPS survey spectra for annealed samples deposited under different oxygen flow conditions. All detected signals can be attributed to Cu, W, O, and C, indicating negligible contamination \cite{Moulder1992XPSHandbook}.

The Cu\,2p region shows the most pronounced changes with oxygen flow and is therefore discussed first. Figures~\ref{fig:survey}(b) and (c) show the Cu\,2p spectra, revealing spin--orbit split peaks corresponding to Cu\,2p$_{3/2}$ and Cu\,2p$_{1/2}$, together with characteristic satellite features of Cu$^{2+}$ \cite{Moulder1992XPSHandbook, Roychowdhury2020Multiinstrument, Biesinger2010XPSStates, Biesinger2017CuXPS}.
Figure~\ref{fig:survey}(b) shows the spectra with a baseline offset, where each signal has been vertically shifted so that its minimum intensity is zero to facilitate a direct comparison of the peak evolution. Under identical measurement conditions, the Cu\,2p peak area is larger for films deposited at 5 and 7.5~sccm, whereas films deposited at higher oxygen flows ($\geq$12.5~sccm) exhibit similar peak areas with only small variations.
Figure~\ref{fig:survey}(c) presents normalized spectra. A systematic shift of the Cu\,2p peaks toward higher binding energies is observed with increasing oxygen flow, while the lineshape and full width at half maximum (FWHM) remain essentially unchanged.

In contrast to Cu\,2p, the W\,4f spectra show no significant changes in peak position, width, or lineshape. Figures~\ref{fig:survey}(d) and (e) display the spin--orbit split W\,4f$_{7/2}$ and W\,4f$_{5/2}$ components \cite{Moulder1992XPSHandbook, Roychowdhury2020Multiinstrument}.
Figure~\ref{fig:survey}(d) indicates that the W\,4f intensity (also displayed with a baseline offset) increases as the oxygen flow rate rises from 5 to 12.5~sccm and remains nearly constant for higher flows—a trend opposite to that observed for Cu\,2p in Fig.~\ref{fig:survey}(b). Furthermore, the normalized W\,4f spectra in Fig.~\ref{fig:survey}(e) confirm the absence of detectable variations in binding energy, FWHM, or overall lineshape across the entire sample series.

Figure~\ref{fig:survey}(f) shows the normalized O\,1s signal \cite{Moulder1992XPSHandbook}. Similar to Cu\,2p, the O\,1s peak shifts toward higher binding energies with increasing oxygen flow.

For clarity, Figures~\ref{fig:survey}(c), (e), and (f) are plotted over a 35~eV binding energy range, facilitating direct comparison of the peak shifts in Cu\,2p, W\,4f, and O\,1s.

\medskip
\noindent\textit{(ii) Auger analysis and Wagner plot.}
\medskip

Figure~\ref{fig:survey}(g) shows the main peak of the Cu LMM Auger spectra (Cu L$_3$M$_{45}$M$_{45}$) for all samples. The spectral shape closely resembles that reported for Cu$^{2+}$ compounds~\cite{Biesinger2017CuXPS}. As observed by comparing Fig.~\ref{fig:survey}(g) and (c), the Cu 2p$_{3/2}$ binding energy and Cu LMM kinetic energy shift in a correlated manner. This indicates that the modified Auger parameter,
\begin{equation}
\alpha' = E_{\text{binding}}(\text{Cu 2p}_{3/2}) + E_{\text{kinetic}}(\text{Cu L}_3\text{M}_{45}\text{M}_{45}),
\label{eq:auger}
\end{equation}
remains essentially constant across all annealed films, where $E_{\text{binding}}(\text{Cu 2p}_{3/2})$ and $E_{\text{kinetic}}(\text{Cu L}_3\text{M}_{45}\text{M}_{45})$ denote the binding energy of the Cu~2p$_{3/2}$ core level and the kinetic energy of the Cu L$_3$M$_{45}$M$_{45}$ Auger transition, respectively.

Using a simplified analysis based on survey spectra (GL(30) line shapes and Shirley background), we obtain an average value of $\alpha' = 1851.67 \pm 0.03$~eV for the samples deposited at different oxygen flows shown in Fig. \ref{fig:survey}. This value is consistent with reported values for solid copper Cu (s) ($\alpha' \approx 1851.24$~eV) and Cu$^{2+}$ compounds such as \ch{CuO}, \ch{CuF2}, and \ch{CuSO4}~\cite{Biesinger2017CuXPS,Moretti_Beck_2019,Moretti_Beck_2022_Model_Extension}.

\begin{table}[h]
\centering
\caption{Mean surface \ch{Cu/W} atomic ratio as a function of oxygen flow rate ($\phi_{\mathrm{O}_2}$), determined by XPS. STD denotes the standard deviation of the mean, and $n$ represents the number of independent samples analyzed for each data point.}
\begin{tabular}{cccc}
\hline
$\phi_{\mathrm{O}_2}$ (sccm)  & Cu/W ratio & STD & $n$. samples \\
\hline
5.0  & 10.98 & 1.9 & 2 \\
7.5  & 6.19  & 0.5 & 2 \\
12.5 & 1.23  & n/a & 1 \\
15.0 & 1.67  & 0.4 & 3 \\
17.5 & 1.84  & 0.8 & 3 \\
20.0 & 2.17  & 0.2 & 2 \\
\hline
\end{tabular}
\label{tab:comp}
\end{table}

Within the semi-empirical framework of Moretti and Beck~\cite{Moretti_Beck_2019,Moretti_Beck_2022_Model_Extension}, $\alpha'$ is primarily governed by final-state effects. Its invariance therefore indicates similar final-state screening in all samples. Since $E_{\text{binding}}(\text{Cu 2p}_{3/2})$ contains both initial- and final-state contributions, the observed binding energy shifts (Fig.~\ref{fig:survey}(c)) are attributed to variations in the initial state, i.e., changes in the local chemical environment of Cu.

Figure~\ref{fig:wagner} presents the Wagner plot, i.e., $E_{\text{kinetic}}(\text{Cu L}_3\allowbreak\text{M}_{45}\allowbreak\text{M}_{45})$ as a function of $E_{\text{binding}}(\text{Cu 2p}_{3/2})$. Red open circles correspond to the present samples, while blue filled circles denote literature data for Cu$^{1+}$ and Cu$^{2+}$ compounds compiled by Moretti and Beck~\cite{Moretti_Beck_2019}. Their compilation is primarily based on the dataset of Biesinger~\cite{Biesinger2017CuXPS}, with additional data for \ch{Cu2S} and \ch{CuS} taken from Perry and Taylor~\cite{PerryTaylor1986_Auger_Cu2S_CuS}, the latter being denoted by an asterisk ($\ast$).

Cu compounds are known to form distinct linear trends in the Wagner plot~\cite{Biesinger2017CuXPS,Moretti_Beck_2019}: Cu$^{2+}$ species follow a line of slope $-1$ (constant $\alpha'$), whereas Cu$^{1+}$ species follow a slope close to $-3$. Both trends converge at metallic Cu. The present data clearly align along the Cu$^{2+}$ trend, indicating that the observed shifts arise from initial-state variations under nearly constant final-state screening. This behavior is consistent with efficient local screening via charge transfer to the Cu 3d states of core-ionized Cu$^{2+}$ species~\cite{Moretti_Beck_2019}.

\medskip
\noindent\textit{(iii) Surface composition and phase interpretation.}
\medskip

Consistent with the trends observed in Figures~\ref{fig:survey}(b) and (d), the Cu-to-W atomic ratio (Cu/W) decreases strongly with increasing oxygen flow at low \ch{O2} flux, dropping from Cu/W $\sim$11 at 5~sccm to values close to unity at 12.5~sccm, and remaining relatively constant at higher flows (Table~\ref{tab:comp}).

Samples deposited at low oxygen flows (5 and 7.5~sccm) exhibit Cu/W ratios exceeding the expected range for \ch{CuWO4} + \ch{Cu3WO6} mixtures, which spans from 1 (pure \ch{CuWO4}) to 3 (pure \ch{Cu3WO6}). This indicates that the excess Cu must form a separate phase. The low optical band gaps (Table~\ref{tab:xrd_bandgap}) and the characteristic Cu$^{2+}$ satellite features (Figure~\ref{fig:survey}(b)) point to the presence of CuO, suggesting the formation of an amorphous CuO phase in these samples.

At intermediate and high oxygen flows, the lower Cu/W ratios are consistent with the coexistence of \ch{CuWO4} and \ch{Cu3WO6}. This interpretation is supported by the optical band gap values (Table~\ref{tab:xrd_bandgap}), which agree with literature values for these ternary compounds~\cite{Benko1982Cu3WO6, Rajeshwar2018CopperOxideReview}.

The compositional trends further support the Wagner plot analysis (Fig.~\ref{fig:wagner}), where the data shift systematically away from \ch{CuO}-like values toward higher binding energies while remaining strictly on the \ch{Cu}$^{2+}$ trend line. While specific Wagner plot coordinates for ternary Cu-W-O phases are not widely reported in the literature, the observed correlation between the Cu/W atomic ratio and this coordinate shift strongly points toward a transition from a binary \ch{CuO} environment toward ternary phases, such as \ch{CuWO4} and \ch{Cu3WO6}. This progression, which aligns with our optical findings, illustrates how the \ch{Cu} initial-state energy is modulated as the local coordination evolves into a ternary oxide framework, all while maintaining the \ch{Cu}$^{2+}$ oxidation state and constant final-state screening.

\medskip
\noindent\textit{(iv) Surface vs. bulk composition.}
\medskip

Figure~\ref{fig1}(a) shows that the discharge current of the W target decreases more rapidly than that of Cu with increasing oxygen flow, indicating that the deposited films become progressively Cu-rich. Consistently, XRD results (Table~\ref{tab:xrd_bandgap}) show the emergence of the Cu-rich \ch{Cu3WO6} phase at higher oxygen flows.

However, this trend contrasts with the surface composition measured by XPS at low oxygen flow, where unusually high Cu/W ratios are observed. This discrepancy can be understood by considering the precursor films prior to annealing.

As shown in Figure~\ref{fig2}, films deposited at 5 and 7.5~sccm contain metallic Cu and \ch{Cu2O}, whereas films deposited at higher oxygen flows are XRD-amorphous and already contain Cu predominantly in the Cu$^{2+}$ state (see detailed high-resolution XPS analysis below).

In the low-oxygen samples, the presence of metallic and Cu$^{+}$ species implies higher Cu mobility. This enhanced mobility provides a plausible mechanism for Cu migration toward the surface or along grain boundaries, leading to the formation of CuO-rich regions. Because XPS probes only the near-surface region, this segregation leads to the enhanced Cu/W ratios observed in Table~\ref{tab:comp}.

In contrast, higher oxygen flows produce precursor films in which Cu is already oxidized to Cu$^{2+}$, reducing its mobility \cite{Lee2018CaDopedCuO}. As a result, Cu remains more uniformly distributed, and the phase evolution follows the CuO–\ch{WO3} phase diagram \cite{Koltsova1999_CuOWO3system}. Within this framework, an increase in the Cu/W ratio leads to the formation of the Cu-rich ternary phase, \ch{Cu3WO6}.

This interpretation is consistent with previous reports of Cu migration and segregation in related systems, including a-IGZO transistors \cite{Lee2018CaDopedCuO}, Cu-doped \ch{WO3} films \cite{Gopalan2007CuWO3}, and dielectric/Cu/dielectric stacks \cite{Zubkins2023WO3CuWO3, PerezLopez2012CuMoO3}.
Several strategies have been proposed to suppress this diffusion, including the use of Cu alloys \cite{Tuo2017CuNiDielectric} or the deposition of dense \ch{WO3} layers \cite{Zubkins2023WO3CuWO3}. Of particular relevance here, CuO layers themselves have also been reported to act as effective Cu diffusion barriers \cite{Lee2018CaDopedCuO}.

Accordingly, increasing the oxygen flow promotes the formation of Cu$^{2+}$ in the precursor films, reduces Cu mobility \cite{Lee2018CaDopedCuO}, and suppresses segregation, resulting in behavior consistent with the CuO/\ch{WO3} phase diagram \cite{Koltsova1999_CuOWO3system}.

\medskip
\noindent\textit{(v) High-resolution peak fitting and chemical state identification.}
\medskip

While the survey-based Wagner plot established the broad chemical trends in comparison to literature standards, high-resolution core-level spectra for Cu 2p, O 1s, and W 4f were subsequently acquired to provide the precision necessary for detailed phase analysis and the peak fitting of distinct chemical environments

\medskip

\begin{figure} 
\includegraphics[width=0.5\textwidth]{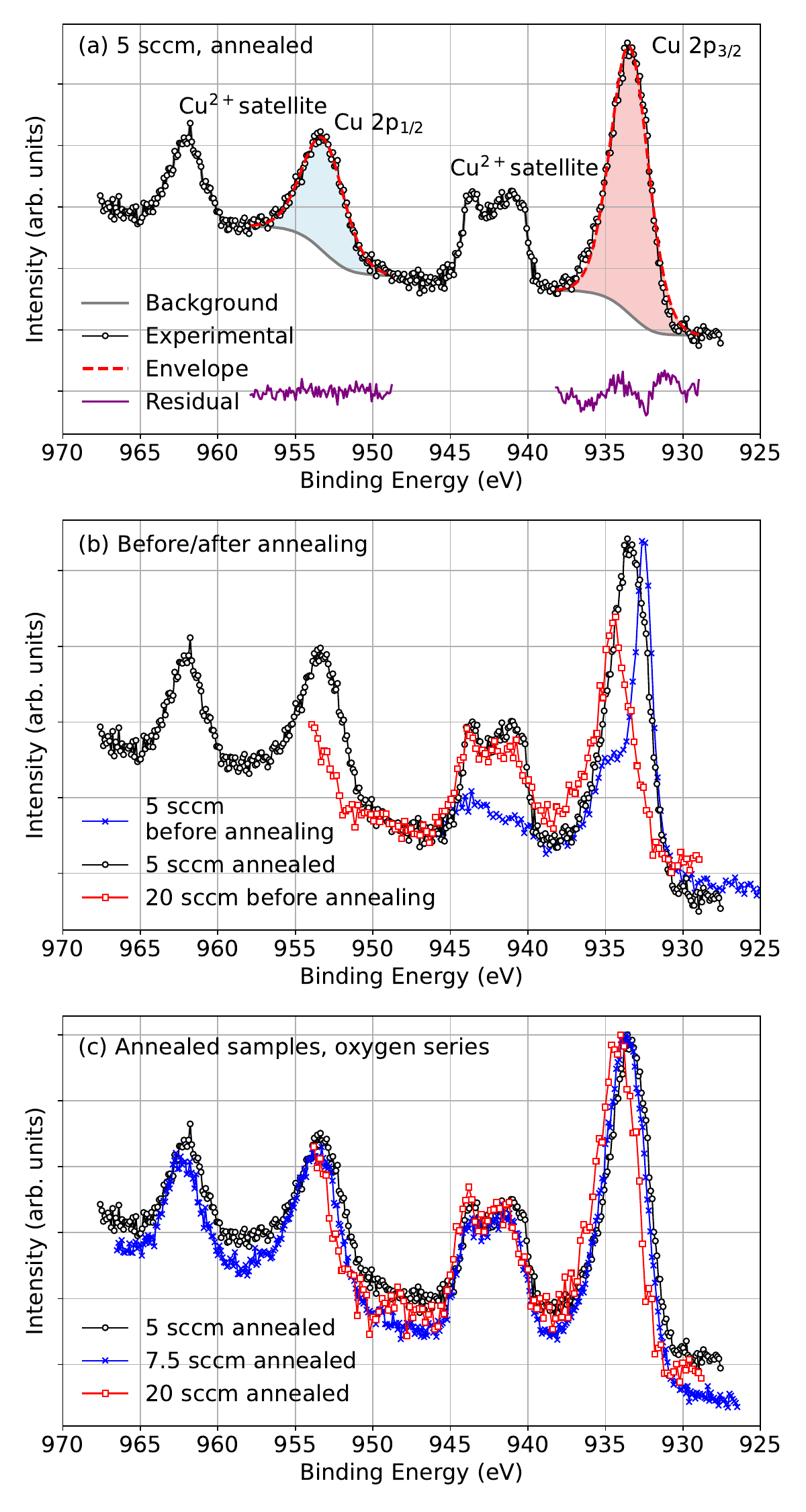}
\caption{Detailed Cu~2p core-level analysis. (a) High-resolution scan and peak fitting of the Cu~2p$_{3/2}$ and Cu~2p$_{1/2}$ doublets for the annealed sample deposited at 5~sccm. (b) Comparison of the Cu~2p region for the 5~sccm sample in the as-deposited and annealed states, including the 20~sccm as-deposited sample for reference. (c) Evolution of the Cu~2p spectra for annealed samples deposited at 5, 7.5, and 20~sccm, illustrating the transition in copper chemical states.}
\label{Cu2p}
\end{figure}

Figure \ref{Cu2p} shows the Cu 2p high-resolution region for samples deposited under different oxygen flow conditions. 
Figure \ref{Cu2p}(a) shows the expected spin–orbit split components, Cu 2p$_{3/2}$ and Cu 2p$_{1/2}$ for a representative 5 sccm annealed sample. These features can be reasonably fitted using a single Gaussian–Lorentzian component for each peak, located at 933.43 eV and 953.18 eV, with FWHM values of 2.89 eV and 3.17 eV, respectively. The measured spin–orbit splitting, $\Delta E$ = -19.75 eV, and the corresponding area ratio of 1/0.46 are in reasonable agreement with the expected values \cite{Biesinger2017CuXPS, Biesinger2010XPSStates}.
The residuals of the fit are also shown in Figure \ref{Cu2p}(a). The standard deviation of the residual is 1.58. A linear fit applied to the background region between peaks yields a standard deviation of 1.45 (not shown), indicating that the residuals are on the order of the experimental noise.
In addition to the main spin–orbit components, Figure \ref{Cu2p}(a) exhibits two pronounced satellite features, located approximately between 940–945 eV and 960–965 eV. The presence of these shake-up satellites is characteristic of Cu$^{2+}$ species, confirming that copper is in its Cu$^{2+}$ oxidation state \cite{Biesinger2017CuXPS, Biesinger2010XPSStates}.

Figure \ref{Cu2p}(b) compares the annealed sample (5 sccm) with unannealed samples deposited at 5 sccm and 20 sccm. Although the spectra for the unannealed samples were acquired over a narrower energy range, the lower-binding-energy shake-up satellite is still captured and serves as a robust diagnostic of the Cu$^{2+}$ electronic structure.
In the unannealed sample deposited at 5 sccm, the Cu 2p$_{3/2}$ peak is noticeably narrower and shifted toward lower binding energies relative to the annealed sample, suggesting the presence of metallic Cu.
In fact, the ISO standard binding energy for metallic Cu$^0$ is 932.62 eV for a monochromatic Al K$\alpha$ X-ray source \cite{seah2001ISOCuXPS}, and exhibits a considerably narrower full width at half maximum (FWHM) than Cu in CuO, in excellent agreement with our observations \cite{Biesinger2010XPSStates}.
A shoulder at higher binding energies is also observed in the Cu 2p$_{3/2}$ peak of the unannealed 5 sccm sample. This feature suggests the presence of an additional copper oxidation state, tentatively assigned to Cu$^{1+}$ in \ch{Cu2O}. The weak satellite intensity further supports this interpretation \cite{Biesinger2017CuXPS, Biesinger2010XPSStates}. Moreover, these observations are consistent with the XRD results presented in Figure \ref{fig2}, where diffraction peaks corresponding to metallic Cu and \ch{Cu2O} were detected in the unannealed 5 sccm sample.
Interestingly, the unannealed sample deposited at 20 sccm, which exhibits an amorphous structure in XRD (Figure \ref{fig2}), displays a pronounced shake-up satellite characteristic of Cu$^{2+}$ species, Figure \ref{Cu2p}(b). This is consistent with the observation anticipated above, and indicates that at higher oxygen flow rates, Cu$^{2+}$ species are already present prior to annealing, whereas lower oxygen flows favor metallic Cu. This supports our hypothesis that oxygen incorporation hinders the diffusion of excess Cu and the subsequent formation of a separate amorphous CuO phase upon annealing; instead, higher oxygen flows stabilize Cu in a less mobile CuO-like state, which promotes the formation of \ch{CuWO4}, while excess Cu leads to \ch{Cu3WO6}.

Figure \ref{Cu2p}(c) shows the Cu 2p region for annealed samples deposited at 5 sccm, 7.5 sccm, and 20 sccm. A systematic shift of the Cu 2p peaks toward higher binding energies is observed ($\Delta E = $+ 1 eV between the 5 and 20 sccm samples). 
To ensure the reliability, repeated measurements were performed on three samples prepared under identical 5 sccm conditions. The resulting average position for the Cu 2p$_{3/2}$ peak was $933.45 \pm 0.10$ eV. This narrow distribution confirms that the 1 eV shift observed at higher oxygen flows is statistically significant and reflects a genuine change in the chemical environment, validating the trends previously noted in the survey-based Wagner plot (Figure \ref{fig:wagner}).
This displacement may be associated with variations in the local electronic environment, likely reflecting the decreasing contribution of CuO and the increasing presence of the \ch{Cu3WO6} phase within the \ch{Cu(WO4}/\ch{Cu3WO6} mixture at higher oxygen flow conditions.

 \begin{figure} 
\includegraphics[width=0.5\textwidth]{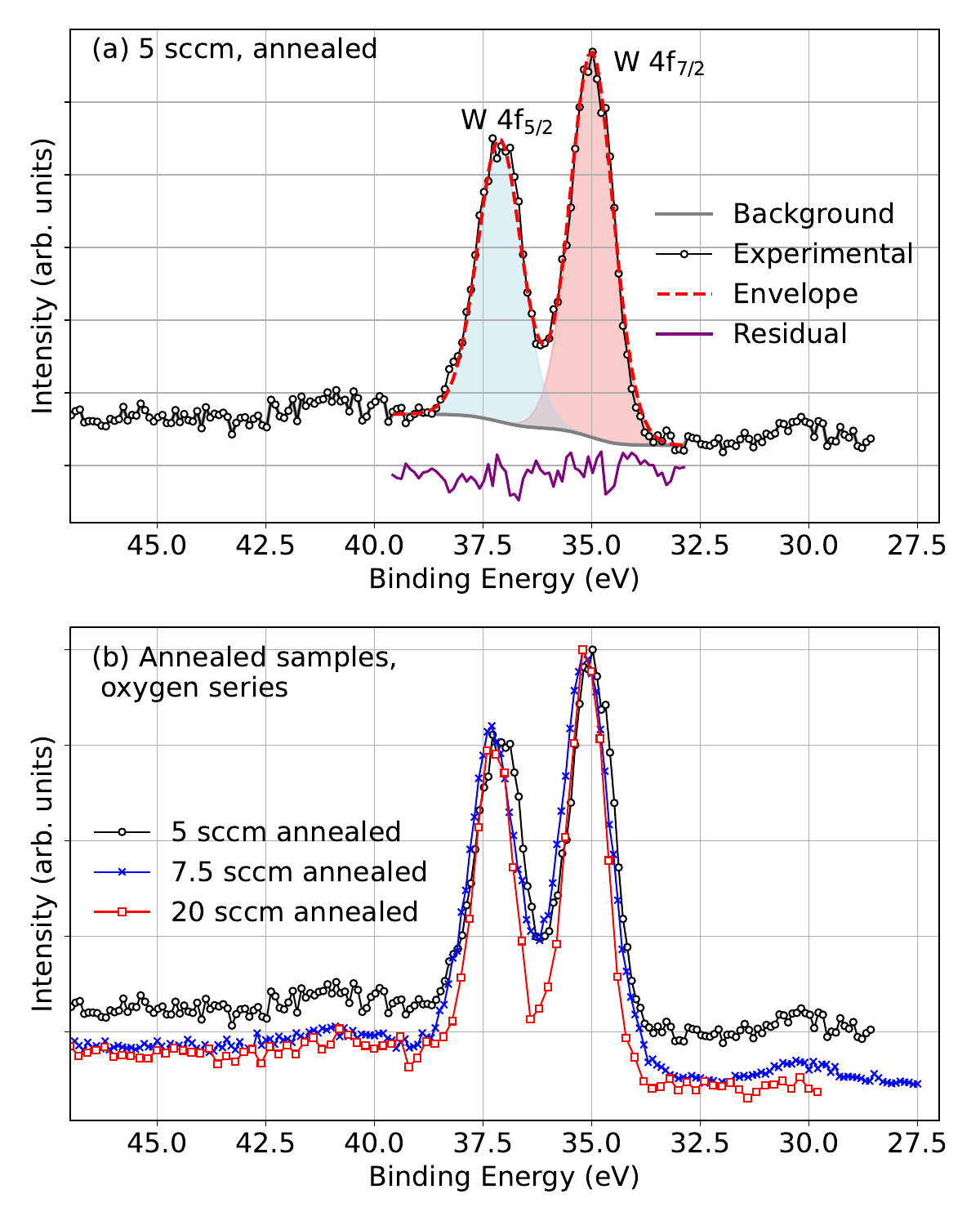}
\caption{Detailed W~4f core-level analysis. (a) High-resolution scan and peak fitting of the W~4f$_{7/2}$ and W~4f$_{5/2}$ doublet for the annealed sample deposited at 5~sccm. (b) Evolution of the W~4f spectra for annealed samples deposited at 5, 7.5, and 20~sccm, illustrating the fair stability of the tungsten chemical state across the flow range.}
\label{W4f}
\end{figure}

\medskip
Figure \ref{W4f}(a) shows the W 4f core-level spectrum for a representative sample deposited at 5 sccm and subsequently annealed. As expected, this binding energy region exhibits the characteristic spin–orbit doublet, consisting of the W 4f$_{7/2}$ and W 4f$_{5/2}$ components. Each component can be satisfactorily fitted using a single Gaussian–Lorentzian line shape. The residual associated with the fitting, displayed in the figure, exhibits a standard deviation of 1.31. For comparison, a linear fit applied to the background in the same energy range, sufficiently removed from the peak positions (not shown), yields a standard deviation of 1.20. The comparable magnitude of these deviations indicates that the residual remains at the background noise level, confirming that each spin–orbit component is adequately described by a single Gaussian–Lorentzian function.
From the peak fitting shown Figure \ref{W4f}(a), the energy separation between the spin–orbit components is determined to be $\Delta E =$ 2.12 eV, with an area ratio of  1/0.75, consistent with the values expected for tungsten in this spectral region \cite{Moulder1992XPSHandbook}. The binding energy position and its standard deviation of the W 4f$_{7/2}$ component is $35.00\pm 0.03$ eV (obtained by comparing three measurements in three samples prepared under identical 5 sccm conditions). The full width at half maximum (FWHM), constrained to be identical for both components, is 1.20 eV. These spectral parameters are consistent with the typical energies reported for W$^{6+}$ \cite{Moulder1992XPSHandbook}. 
Components consistent with W$^{6+}$ are also observed in the spectra acquired prior to the annealing process (not shown). The only exception is the sample deposited at the lowest oxygen flow (5 sccm), where weak features suggestive of a metallic contribution are detected.

Fig. \ref{W4f} (b) presents the W 4f spectra for annealed samples deposited under varying oxygen flow conditions (5, 7.5, and 20 sccm). The W 4f binding energies remain essentially constant across all samples, with a total standard deviation of only 0.06 eV. This value is nearly an order of magnitude smaller than the shift observed in the Cu 2p region and approaches the limit of experimental error. These results indicate that tungsten remains securely in the W$^{6+}$ oxidation state and that its local chemical environment is highly stable, regardless of the oxygen flow rate.

 \begin{figure} 
\includegraphics[width=0.5\textwidth]{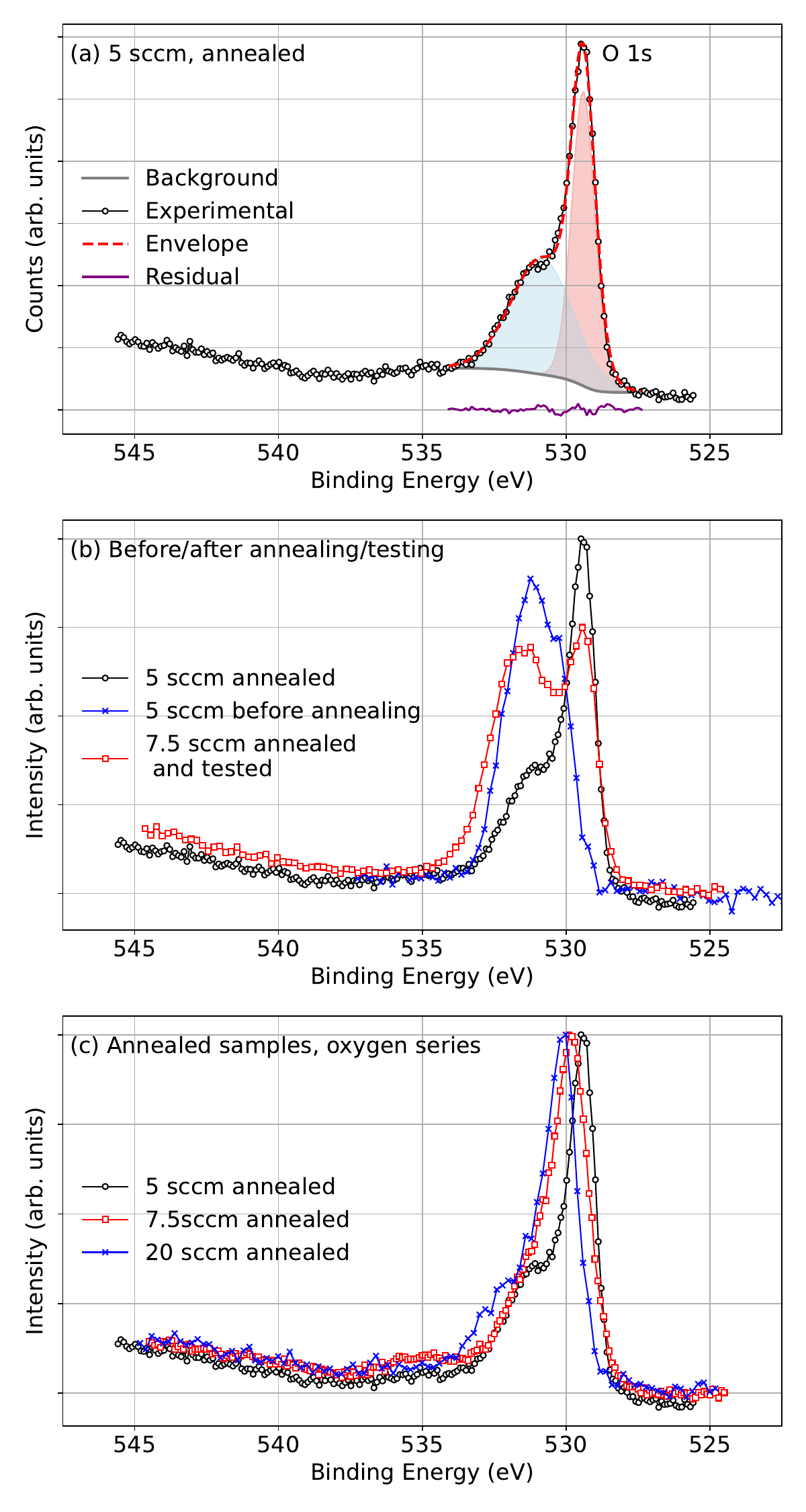}
\caption{Detailed O~1s core-level analysis. (a) High-resolution scan and peak fitting of the O~1s signal for the annealed sample deposited at 5~sccm. (b) Comparison of the O~1s region for the 5~sccm sample in the as-deposited and annealed states, including a 7.5~sccm annealed sample after photocatalytic testing in aqueous solution for reference. (c) Evolution of the O~1s spectra for annealed samples deposited at 5, 7.5, and 20~sccm, illustrating the transition in the oxygen chemical states.}
\label{O1s}
\end{figure}

\medskip
The O 1s region is presented in Figure \ref{O1s}(a) for a representative sample deposited at 5 sccm. The spectrum is satisfactorily fitted using two GL(30) components located at 529.36 eV and 530.92 eV, with full widths at half maximum (FWHM) of 0.99 eV and 2.24 eV, respectively.
The lower binding-energy component, characterized by its narrower linewidth, is assigned to lattice oxygen \cite{Moulder1992XPSHandbook, Svintsitskiy2011CuO_oxide_surface} associated in our case with the \ch{CuWO4} and/or \ch{Cu3WO6} phases (and possibly to other phases including Cu-O compounds )\cite{Svintsitskiy2011CuO_oxide_surface,Tang2016CuWO4}. Crystallographic data reported in ICDD cards 01-080-5325 (\ch{CuWO4}) and 04-009-6702 (\ch{Cu3WO6}) indicate multiple oxygen chemical environments within both structures. In \ch{CuWO4}, oxygen occupies four distinct crystallographic sites, whereas in \ch{Cu3WO6}, two nonequivalent oxygen environments are present. However, the binding-energy separations among these sites are expected to be smaller than the experimental resolution of the XPS measurements. Consequently, the contributions from these distinct environments are not spectroscopically resolved and are collectively represented by a single component attributed to lattice oxygen.
The higher binding-energy component in O1s exhibits a substantially broader FWHM of 2.24 eV. While this feature can be generally associated with non-lattice oxygen environments, its pronounced width indicates that it cannot be attributed to a single, well-defined chemical state. Rather, this component most likely comprises overlapping contributions arising from surface hydroxyl species, adsorbed oxygen-containing groups, and oxygen associated with non-stoichiometric or structurally disordered environments \cite{Moulder1992XPSHandbook,Li2017CuPhotocat,Tang2016CuWO4,Svintsitskiy2011CuO_oxide_surface,PinonEspitia2023CuO_surface_oxygen}.

This interpretation is consistent with the spectra shown in Figure \ref{O1s}(b), which compares the sample deposited at 5 sccm before and after annealing, together with an annealed sample following photocatalytic testing (7.5 sccm) performed in an aqueous solution. Although the photocatalytic experiment is not part of the present study, the tested sample is included here for comparison, as it provides useful insight into the evolution of surface species. In both cases — namely, the as-deposited sample and the sample exposed to aqueous photocatalytic conditions — the relative intensity of the higher binding-energy component increases significantly, eventually becoming dominant. The increase in the annealed sample following photocatalytic testing is commonly reported in Cu oxide \cite{Li2017CuPhotocat} and tungstate systems \cite{Tang2016CuWO4} and is typically attributed to enhanced surface hydroxylation and adsorption processes induced by ambient exposure or interaction with aqueous media, whereas in the as-deposited sample, the feature reflects greater structural disorder in the unannealed material, which also contributes to this component.

The fitting procedure employing two GL components, as shown in Figure \ref{O1s}(a), yields a residual with a standard deviation of 1.53, comparable to that obtained for the background fitted using a linear function (1.45). This agreement indicates that the selected fitting model adequately reproduces the experimental spectrum without introducing unnecessary parameters.
While adding further components would inevitably reduce the residuals, such an approach would introduce solutions lacking a solid physical interpretation. The higher binding-energy feature encompasses contributions from chemically similar species—including hydroxyl groups, adsorbed oxygen, and oxygen associated with non-stoichiometric or structurally disordered environments—whose binding-energy separations fall below reliable spectral discrimination limits. Consequently, decomposing this signal into multiple discrete peaks would result in over-parameterization without providing robust chemical significance. The broader FWHM of this component is therefore interpreted as a physically meaningful representation of these unresolved contributions, rather than evidence for spectroscopically separable states.

Consistent with the behavior observed in the Cu 2p region, but in contrast to the W 4f spectra, the O 1s peak position exhibits a systematic shift toward higher binding energies with increasing oxygen flow during deposition. This trend is illustrated in Figure \ref{O1s}(c), where the O1s spectra of annealed samples prepared at 5, 7.5, and 20 sccm are compared. 
The position and standard deviation of the main lattice-oxygen component, based on three separate measurements, is $529.36 \pm 0.03$ eV for the 5 sccm sample. The observed shift of $\Delta\mathrm{E}=0.38$ eV between the samples deposited at 5 and 20 sccm samples is more than ten times the measured standard deviation, confirming that the shift is statistically significant.

\medskip
The quantified shifts from the high-resolution regions confirm the general trends observed with survey data and provide a clear narrative of the phase evolution: Cu acts as the primary active site, experiencing the most dramatic change in its local electronic environment as the system transitions toward higher-oxygen content phases. The oxygen is partially affected as its bonding environment reflects this transition, while the tungsten remains in a highly stable configuration. It is worth noting here that the Cu data follow a slope of -1 on the Wagner plot (Figure \ref{fig:wagner}), indicating a constant modified Auger parameter. This demonstrates that the observed shift is governed by initial-state effects--specifically a modification of the Cu ground--state electronic structure and Cu-O-W hybridization, rather than changes in final-state relaxation or screening.

\section*{Conclusions}

In this study, Cu-W-O thin films were synthesized via reactive DC magnetron co-sputtering followed by thermal annealing at $500~^\circ\mathrm{C}$. Our findings demonstrate that the phase evolution and surface chemistry of the Cu–W–O system are critically governed by the oxygen flow rate during deposition.

At higher oxygen flows, XRD confirms the coexistence of \ch{CuWO4} and a Cu-rich  \ch{Cu3WO6} cubic phase. In this regime, the \ch{CuWO4} phase exhibits a polycrystalline structure that aligns with the ICDD standard. Conversely, films deposited at low oxygen flows nominally suggest a pure triclinic \ch{CuWO4} phase, again according with the ICDD standard, but with a distinct (0-21) preferential orientation. 
Despite the XRD-apparent purity of these low-oxygen samples, optical characterization reveals a reduced bandgap indicative of an undetected amorphous \ch{CuO} phase. Wagner plot analysis corroborates this finding: at low oxygen flows, the sample coordinates align with the characteristic region for \ch{CuO}, providing a chemical signature for the amorphous phase that remains undetected by diffraction.

Furthermore, we propose that increased oxygen availability during sputtering effectively immobilizes Cu atoms within the precursor matrix, thereby suppressing long-range diffusion and surface segregation. This interpretation is supported by the discrepancy between sputtering discharge current trends, XRD data and the measured surface composition. While the pronounced decrease in W discharge current at high oxygen flows would be expected to favor a Cu-enriched flux, XPS instead shows the highest surface Cu concentration in low-oxygen samples. Likewise, although XRD identifies the Cu-rich \ch{Cu3WO6} phase at high oxygen flows, surface analysis reveals comparatively lower Cu content under these conditions.
This phenomenological discrepancy--where both discharge current and XRD suggest Cu enrichment at high oxygen flows, yet surface analysis indicates the opposite--implies that elevated oxygen partial pressure during deposition stabilizes Cu within the bulk. As a result, surface segregation and the formation of independent amorphous \ch{CuO}, observed at low oxygen flows, are suppressed.

Moreover, electronic characterization confirms that while the W chemical environment remains  stable across all conditions, the Cu chemical environment is highly sensitive to processing parameters. The binding energy shift observed in the \ch{Cu} 2p$_{3/2}$ transition, verified by the constant modified Auger parameter, is dominated by initial-state effects. Given that surface chemical states dictate functional performance in catalysis and sensing, these results highlight that materials loosely labeled as \ch{CuWO4} in the literature may in fact span a wide spectrum of chemical states. Ultimately, the additional degrees of freedom introduced by the second cation necessitate careful chemical validation to disentangle intrinsic ternary behavior from synthesis-induced heterogeneity. In this work, we show that systematic Wagner plot analysis provides a reliable benchmark for validation and a straightforward way to track the evolution of Cu–W–O surfaces while ensuring reproducibility across different synthesis methods.

\section*{Acknowledgements}

The author acknowledges financial support from the Beatriz Galindo Senior Grant 2024 for the attraction of talent, funded by the Spanish Ministerio de Ciencia, Innovación y Universidades.
The author would like to thank Prof. Tomas Edvinsson (Uppsala University) for hosting a research stay at the \AA ngstr\"om Laboratory (Uppsala University), which made this work possible.
%

\FloatBarrier

\bibliography{bibliography}
\bibliographystyle{elsarticle-num}

\end{document}